\newcommand{\paperOption}{12pt,onecolumn}
\newcommand{\usehyperref}{%
 \usepackage[pdfstartview=FitH,hidelinks]{hyperref}%
}
\newcommand{\copyrightnotice}{%
\copyright\ 2021 IEEE. Personal use of this material is permitted.
Permission from IEEE must be obtained for all other uses, in any current or future media, including reprinting/republishing this material for advertising or promotional purposes, creating new collective works, for resale or redistribution to servers or lists, or reuse of any copyrighted component of this work in other works.%
}
\InputIfFileExists{LAB.opt}{%
 \typeout{Options overridden!}%
}{}

\documentclass[\paperOption]{IEEEtran}
\usepackage{amsmath,amssymb,amsthm}
\usepackage{dsfont}
\usepackage{algorithmicx}
\usepackage{algorithm}
\usepackage{algpseudocode}
\usepackage{nccmath}
\usepackage{bm}
\usepackage{color}
\usepackage[noadjust]{cite}
\usepackage{stfloats}
\usepackage{float}
\usepackage{multicol}
\usehyperref
\usepackage{multirow}
\usepackage{url}
\usepackage{enumerate}
\usepackage{graphicx}

\theoremstyle{definition}
\newtheorem{definition}{Definition}
\newtheorem{theorem}{Theorem}
\newtheorem{lem}{Lemma}

\newtheorem{prop}{Proposition}

\theoremstyle{remark}
\newtheorem{remark}{Remark}

\algnewcommand{\algorithmicand}{\textbf{ and }}
\algnewcommand{\algorithmicor}{\textbf{ or }}
\algnewcommand{\OR}{\algorithmicor}
\algnewcommand{\AND}{\algorithmicand}

\usepackage{mathenv}

\newcommand{\borel}{\mathcal{B}}

\newcommand{\eqdef}{\triangleq}
\DeclareMathOperator{\expect}{\mathbb{E}}
\newcommand{\pinteger}{\mathds{N}}
\newcommand{\real}{\mathds{R}}
 \newcommand{\nnreal}{\real_{\ge 0}}
 \newcommand{\preal}{\real_{>0}}
\newcommand{\relvar}[2]{\buildrel \mathrm{#2} \over #1}
 \newcommand{\eqvar}[1]{\relvar{=}{#1}}
 \newcommand{\levar}[1]{\relvar{\le}{#1}}
 \newcommand{\ltvar}[1]{\relvar{<}{#1}}
\DeclareMathOperator{\mcr}{MCR}

\title{On Optimal Power Control for Energy Harvesting Communications with Lookahead}

\author{%
 Ali~Zibaeenejad,
 Shengtian~Yang,~\IEEEmembership{Senior~Member,~IEEE,}
 and Jun~Chen,~\IEEEmembership{Senior~Member,~IEEE}%
 \thanks{%
  This work was supported in part by the National Natural Science Foundation of China under Grant 62171411 and in part by the Natural Sciences and Engineering Research Council (NSERC) of Canada under a Discovery Grant.
  This paper was presented in part at the IEEE International Symposium on Information Theory, Paris, France, July 2019 \cite{zibaeenejad_optimal_2019}.
  (Corresponding authors: Ali~Zibaeenejad and Shengtian~Yang.)}
 \thanks{%
  A.~Zibaeenejad and J.~Chen are with the Department of Electrical and Computer Engineering, McMaster University, Hamilton, ON L8S 4K1, Canada (e-mail: \{azibaeen, chenjun\}@mcmaster.ca).
  A.~Zibaeenejad is also with the School of Electrical and Computer Engineering, Shiraz University, Shiraz, Fars 71348-51154, Iran (e-mail: zibaeenejad@shirazu.ac.ir).}
 \thanks{%
  S.~Yang is with the School of Information and Electronic Engineering (Sussex Artificial Intelligence Institute), Zhejiang Gongshang University, Hangzhou 310018, China (e-mail: \mbox{yangst@codlab.net}).}
 \thanks\copyrightnotice
}

\begin{document}

\maketitle

\begin{abstract}
Consider the problem of power control for an energy harvesting communication system, where the transmitter
is equipped with a finite-sized rechargeable battery and is
able to look ahead to observe a fixed number of future energy arrivals. An implicit characterization of the maximum average
throughput over an additive white Gaussian noise channel and the associated optimal power control policy is provided via the Bellman equation under the assumption that the energy arrival process is stationary and memoryless. A more explicit characterization is obtained for the case of Bernoulli energy arrivals by means of asymptotically tight upper and lower bounds on both the maximum average throughput and the optimal power control policy. Apart from their pivotal role in deriving the desired analytical results, such bounds are highly valuable from a numerical perspective as they can be efficiently computed using convex optimization solvers.
\end{abstract}


\section{Introduction}\label{introduction}

Supplying required energy of a communication system
by energy harvesting (EH) from natural energy resources
is not only beneficial from an environmental standpoint,
but also essential for long-lasting self-sustainable affordable
telecommunication which can be deployed in places with no
electricity infrastructure. On the other hand, the EH systems
need to handle related challenges, such as varying nature of
green energy resources and limited battery storage capacity, by employing suitable power control policies \cite{SMJG10, OTYUY11, YU12, TY12, RM12, HZ12, ulukus, BGD13, WL13, SK13, AMM13, XZ14, RSV14, VJ14, UYESZGH15, DFO15, KO15, ozgur16, ABU18, ZP18, ZP182, WZJC19, WZJCarXiv, FYY19, YC20}.
Roughly speaking, the objective of a power control policy for EH communications is to specify the energy assignment across the time horizon based on the available information about the energy arrival process as well as the battery capacity constraint to maximize the average throughput (or other rewards). There are two major categories of power control policies: online and offline.

For online power control, it is assumed that the energy
arrivals are causally known at the transmitter (TX).
Reference \cite{ozgur16} presents the
optimal online power control policy for Bernoulli energy arrivals and establishes the
approximate optimality of the fixed fraction policy for general
energy arrivals. Similar results are derived in \cite{ABU18} for a
general concave and monotonically increasing reward function. Other notable results in this direction include the optimality of the greedy policy in the low battery-capacity regime  \cite{WZJC19,WZJCarXiv} and the characterization of a 
maximin optimal online power control policy \cite{YC20}. 

For offline power control, the TX is assumed to know the realization of the whole energy arrival process in advance. The optimal offline power control policy for the classical  additive white Gaussian noise (AWGN) channel is derived in  \cite{ulukus}. The
analyses of the offline model for more general fading channels
can be found in \cite{OTYUY11,HZ12} (and the references therein). In general,
the optimal policies for the offline model strive to allocate the energy across the time horizon as uniformly as possible while
trying to avoid energy loss due to battery overflow.

From a practical perspective, the offline model is over-optimistic whereas the online model is over-pessimistic.
In reality, the TX often has a good idea
of the amount of available energy in near future, either because such
energy is already harvested but not yet converted to a usable
form%
\footnote{
An energy harvesting module usually contains a storage element such as a supercapacitor or a rechargeable battery.
Under certain EH-module designs, the charging time of the storage element may result in a considerable delay before the harvested energy is ready for use.
For example, a 100 F supercapacitor with an internal DC resistance of 10 m$\Omega$ (see e.g., \cite{maxwell}) has a time constant of $0.01\cdot 100 = 1$ s, so it will take a few seconds to fully charge the supercapacitor.
}
or because the nature of energy source renders it possible to make accurate short-term
predictions%
\footnote{
Predictions can be made based on either past energy arrivals \cite{KHZS07, MKLK19, MKK20} or certain side information.
The latter scenarios arise in energy harvesting from controllable energy sources (e.g., wireless power transfer), where the energy harvester may be informed of the upcoming energy delivery schedule or may even infer the schedule from related information available under a certain protocol of energy management.
This kind of side information is also quite common for weather forecast, which can be leveraged for solar energy estimation. For example, if location A is ahead of location B by a certain amount of time in terms of weather conditions, then this time difference provides a lookahead window for location B. It is worth noting that side-information-based predictions might be feasible even when the energy arrival process is memoryless.}.
From a theoretical perspective, these two models have been largely studied in isolation, and it is unclear how the respective results are related. For example, the distribution of the energy arrival process is irrelevant to the characterization of the optimal control policy for the offline model, but is essential for the online model; moreover, there is rarely any comparison between the two models in terms of the maximum average throughput except for some asymptotic regimes.

In this paper, we study a new setup where the TX is able to
look ahead to observe a window of size $w$ of future energy
arrivals. Note that the online model and offline model correspond
to the extreme cases $w=0$ and $w=\infty$, respectively.
Therefore, our formulation provides a natural link between these two
models. It also better approximates many real-world scenarios.
On the other hand, the new setup poses significant technical challenges. Indeed, it typically has a much larger state space as compared to the online model, and requires an extra effort to deal with the stochasticity of the energy arrival process in the characterization of the optimal power control policy as compared to the offline model. Some progress is made in this work towards overcoming these challenges.
In particular, we are able to provide a complete solution for the case of Bernoulli energy arrivals, depicting a rather comprehensive picture with the known results for the offline model and the online model at two ends of a spectrum.
The finite-dimensional approximation approach developed for this purpose is of theoretical interest in its own right and has desirable numerical properties.

The rest of this paper is organized as follows. We introduce the system model and the problem formulation in Section \ref{Sec:Problem-Def}. Section \ref{Sec:Bellman} provides a characterization of the maximum average throughput via the Bellman equation and a sufficient condition for the existence of an optimal stationary policy. More explicit results are obtained for the case of Bernoulli energy arrivals.
We summarize these results in Section~\ref{Sec:Strategy} and give a detailed exposition in Section \ref{sec:proof} with proofs relegated to the appendices. The numerical results are presented in Section \ref{sec:numerical}. Section \ref{Sec:Conclusion} contains some concluding remarks.

Notation: $\mathds{N}$ and $\real$ represent the set of positive integers and the field of real numbers, respectively.
Random variables are denoted by capital letters and their realizations are written in lower-case letters.
Vectors and functions are represented using the bold font and the calligraphic font, respectively.
An $N$-tuple $(x_1, x_2, \ldots, x_N)$ is abbreviated as $(x_j)_{j=1}^N$.
Symbol $\mathbb{E}$ is reserved for the expectation.
The logarithms are in base~2.

\section{Problem Definitions} \label{Sec:Problem-Def}

The EH communication system considered in this paper consists of a TX with EH capability, a receiver (RX), and a connecting point-to-point quasi-static fading AWGN channel.
The system operates over discrete time slots $\tau \in \mathds{N}$ as formulated by
\begin{equation} \label{eq:channel-model}
  Y_{\tau} = \sqrt{\gamma} X_{\tau} + Z_{\tau}.
\end{equation}
Here the channel gain $\gamma \in \preal$ is assumed to be constant for the entire communication session; $X_{\tau}$ and $Y_{\tau}$ are the transmitted signal and the received signal, respectively; $Z_{\tau}$ is the random white Gaussian noise with zero mean and unit variance. The TX is equipped with a rechargeable battery of finite size $B>0$, which supplies energy for data transmission. In this paper, the TX follows the harvest-save-transmit model: it harvests energy from an exogenous source, stores the harvested energy in the battery (which is limited by the battery capacity $B$), and assigns the stored energy to the transmitted signal according to a pre-designed power control policy.
The harvested energy arrivals, denoted by $\{E_\tau, \tau \in \mathds{N}\}$,  are assumed to be  i.i.d.  with marginal distribution $P_E$.
The variation of the battery energy level over time can be expressed as a random process $\{B_\tau, \tau \in \mathds{N}\}$ with $B_1\eqdef \min\{\beta+E_1,B\}$, where $\beta \in [0, B]$ denotes the initial energy level of the battery.
Based on the harvest-save-transmit model, the TX is able to emit any energy $A_\tau$ in time slot $\tau$ as long as the \emph{energy causality} condition~\cite{ulukus}
\begin{equation} \label{eq:energy-Causality}
  0\leq A_\tau \leq B_{\tau}
\end{equation}
is met, then the battery level is updated to
\begin{equation} \label{eq:battery-change-relation}
  B_{\tau+1} = \min\{B_{\tau} - A_{\tau} + E_{\tau+1} , B\}.\\
\end{equation}

This paper considers a new EH communication system model as follows.
\begin{definition}
An EH communication system is said to have the ``lookahead" capability if the TX  is able to observe the realization of the future energy arrivals within a lookahead window of size $w\in \mathds{N}$ from the transmission time. Specifically, at any transmission time $\tau\in \mathds{N}$, the harvested energy sequence $(E_t)_{t=1}^{\tau+w}$ is known at the TX.
\end{definition}

A power control policy for an EH communication system with lookahead is a
sequence of mappings from the observed energy arrivals to a nonnegative
action value (instantaneous transmission energy). Its precise definition is given below.
\begin{definition}
For an EH communication system with initial battery energy level $\beta$ and lookahead window size $w$, its power control policy, denoted by $\pi_{\beta}(w)$,  is a sequence of  functions $(\mathcal{A}_\tau)_{\tau=1}^\infty$, where  $\mathcal{A}_\tau: [0,B] \times \nnreal^{\tau+w} \rightarrow [0,B]$ specifies the amount of energy used in time slot $\tau$
\begin{equation}\label{def:action-function-general}
  A_\tau = \mathcal{A}_\tau(\beta, E_1, E_2, \ldots, E_{\tau+w})
\end{equation}
subject to conditions~\eqref{eq:energy-Causality} and~\eqref{eq:battery-change-relation}.
\end{definition}
\indent With the consumption of energy $A_{\tau}$, the EH communication system is assumed to achieve the instantaneous rate given by the capacity~\cite{cover} of channel~\eqref{eq:channel-model}
\begin{equation}\label{eq:reward-per-slot}
  \mathcal{R}(A_\tau) = \frac{1}{2}\log(1+ \gamma A_\tau)
\end{equation}
as the \emph{reward} at time $\tau$. For a fixed communication session $T \in \mathds{N}$,  the $T$-horizon expected throughput induced by  policy $\pi_{\beta}(w)$ is defined as 
\begin{equation} \label{eq:average-throughput-finite}
\Gamma^{\pi_{\beta}(w)}_T \triangleq \frac{1}{T} \mathbb{E}\left(\sum\limits_{\tau=1}^T \mathcal{R}(A_\tau)\right),
\end{equation}
where the expectation is over the energy arrival sequence $(E_t)_{t=1}^{T}$.
\begin{definition} \label{def:long-term-throughput}
	We say that $\pi^*_{\beta}(w)$ is an optimal power control policy if it achieves  the maximum (long-term) average throughput 
\begin{equation} \label{eq:long-term-throughput}
  \Gamma^*_{\beta} \triangleq \sup_{\pi_{\beta}(w)} \liminf_{T\rightarrow \infty} \Gamma^{\pi_{\beta}(w)}_T.
\end{equation}
\end{definition}

\begin{remark}
By an argument similar to \cite[Prop.~6]{ozgur16}, it can be shown that $\Gamma_\beta^*$ does not depend on $\beta$.
Hence, the subscript $\beta$ in \eqref{eq:long-term-throughput} can be dropped, and we can assume $\beta=B$ without loss of generality.
\end{remark}

\section{Bellman Equation and the Existence of an Optimal Stationary Policy} \label{Sec:Bellman}

A power control policy $\pi(w)$ is said to be Markovian if each function $\mathcal{A}_\tau$ of  $\pi(w)$ depends on the history $(\beta, E_1, E_2, \ldots, E_{\tau+w})$ only through the current battery state $B_\tau$ and the energy arrivals in the lookahead window $(E_{\tau+1}, E_{\tau+2}, \ldots, E_{\tau+w})$.
If furthermore $\mathcal{A}_\tau$ is time-invariant and deterministic, we say $\pi(w)$ is (deterministic Markovian) stationary.
In this case, $\pi(w)$ can be simply identified by a deterministic mapping $\mathcal{A}$ from the state space
\[
S\eqdef [0,B]\times \nnreal^w
\]
to $[0,B]$ satisfying $\mathcal{A}(b,e_1,e_2,\ldots,e_w)\le b$.

By \cite[Th.~6.1]{Survey}, if there is a stationary policy $\mathcal{A}^*$ satisfying the so-called Bellman equation \eqref{eq:bellman}, then $\mathcal{A}^*$ is optimal in the sense of \eqref{eq:long-term-throughput}.
\begin{theorem}\label{thm:Bellman}
If there is a constant $g$ and a real-valued bounded function $\mathcal{H}$ on $S$ such that
\begin{eqnarray}
g + \mathcal{H}(b,e_1,e_2,\ldots,e_w)
&= &\sup_{a\in [0,b]} (\mathcal{R}(a) \nonumber\\
\multicolumn{3}{l}{\quad + \expect \mathcal{H}(\min\{b-a+e_1,B\},e_2,e_3,\ldots,e_w,E))}\label{eq:bellman}
\end{eqnarray}
for all $(b,e_1,e_2\ldots,e_w)\in S$, then $\Gamma^*=g$, where $E$ is a nonnegative random variable with distribution $P_E$.
Furthermore, if there exists a stationary policy $\mathcal{A}^*: S\to [0,B]$ such that
\begin{eqnarray*}
\multicolumn{3}{l}{g + \mathcal{H}(b,e_1,e_2,\ldots,e_w)}\\
\quad &= &\mathcal{R}(a)+\expect \mathcal{H}(\min\{b-a+e_1,B\},e_2,e_3,\ldots,e_w,E)
\end{eqnarray*}
with $a=\mathcal{A}^*(b,e_1,e_2,\ldots,e_w)$ for all $(b,e_1,e_2\ldots,e_w)\in S$, then $\mathcal{A}^*$ is optimal.
\end{theorem}

The Bellman equation~\eqref{eq:bellman} rarely admits a closed-form solution, and consequently is often analytically intractable.
Nevertheless, it can be leveraged to numerically compute the maximum average throughput and the associated optimal stationary policy. On the other hand, the numerical solution to~\eqref{eq:bellman} typically does not offer a definite conclusion regarding the existence of an optimal stationary policy. Fortunately, the following result provides an affirmative answer under a mild condition on $P_E$. It will be seen that knowing the existence of an optimal stationary policy sometimes enables one to circumvent the Bellman equation by opening the door to other  approaches.

\begin{theorem}\label{th:StationaryPolicyOptimality}
If $P_E\{E\ge B\}>0$, then there exists an optimal stationary policy achieving the maximum average throughput, regardless of the initial battery state $\beta$.
\end{theorem}

\begin{IEEEproof}
Because of the limit imposed by battery capacity $B$, we assume with no loss of generality that $P_E=P_{\min\{E,B\}}$, and consequently $P_E((B,+\infty))=0$.

Let $p\eqdef P_E([B,+\infty)) = P_E(B)$ and
\[
\mathfrak{f}(A)
\eqdef \frac{p^{w+1}}{B^{w+1}}\mathfrak{m}(A\cap [0,B]^{w+1}) + \delta_{\mathbf{B}^{w+1}}(A),\quad A\in\borel(S),
\]
where $\mathfrak{m}$ denotes the Lebesgue measure on $(S,\borel(S))$, $\delta_x(A)\eqdef 1\{x\in A\}$, $\mathbf{B}^{w+1}$ denotes the $(w+1)$-dimensional all-$B$ vector, and $\borel(S)$ denotes the Borel $\sigma$-algebra of $S$.
Let $\epsilon = p^{w+1}$.
It is clear that any (measurable) set $D$ with $\mathfrak{f}(D)
\le \epsilon$ does not contain $\mathbf{B}^{w+1}$.
Consequently,
\begin{eqnarray*}
P_{S_{w+2}\mid S_1}(D\mid s)
&\le &P_{S_{w+2}\mid S_1}(S\setminus\{\mathbf{B}^{w+1}\}\mid s)\\
&\le &1-p^{w+1}
= 1-\epsilon
\end{eqnarray*}
for all $s\in S$ and all (admissible) randomized stationary policies, where
\[
S_\tau\eqdef (B_{\tau},E_{\tau+1},\ldots,E_{\tau+w}).
\]
This means that the Doeblin condition is satisfied, and hence there exist a set $C\in\borel(S)$ with $\mathfrak{f}(C)>\epsilon$ and a stationary policy $\mathcal{A}^*$ such that for all $S_1=s\in C$, policy $\mathcal{A}^*$ achieves the maximum average throughput and $P_{S_2\mid S_1}(C\mid s) = 1$ (\cite[Th.~2.2]{kurano_existence_1989}).

It is also easy to see that $\mathbf{B}^{w+1}\in C$ and it is accessible (with probability $\ge p^{w+1}$) from any state $s\in S$. Therefore, the Markov chain under $\mathcal{A}^*$ is a unichain, consisting of the recurrent class $C$ and the set $S\setminus C$ of transient states, all yielding the same long-term average throughput.
\end{IEEEproof}

\section{Maximum Average Throughput for Bernoulli Energy Arrivals}\label{Sec:Strategy}

Even with Theorems~\ref{thm:Bellman} and~\ref{th:StationaryPolicyOptimality} at our disposal, it is still very difficult, if not impossible, to obtain an explicit characterization of the maximum average throughput and the associated optimal policy for a generic energy arrival process. To make the problem more tractable, in the rest of this paper we will focus on a special case: Bernoulli energy arrivals.
In this case, the energy arrivals can only take two values, $B$ with probability $p$ and $0$ with probability $1-p$, where $p\in (0,1)$.
Formally, the probability distribution of energy arrivals is given by
\begin{equation}\label{eq:EH-Seq-distribtion}
P_E(A)
\eqdef p\delta_B(A) + (1-p)\delta_0(A),
\end{equation}
where $\delta_x(A)\eqdef 1\{x\in A\}$.

As defined in Section~\ref{Sec:Bellman}, a stationary policy\footnote{In light of Theorem \ref{th:StationaryPolicyOptimality}, it suffices to consider stationary policies.} $\mathcal{A}$ determines the action based on the current system state
\[
S_\tau
\eqdef (B_{\tau},E_{\tau+1},\ldots,E_{\tau+w}).
\]
Let $D_\tau\eqdef \mathcal{D}((E_{\tau+i})_{i=1}^w)$, where
\begin{eqnarray*}
\multicolumn{3}{l}{\mathcal{D}((e_i)_{i=1}^w)}\\
\quad &\eqdef &\max\left\{1\le i\le w+1: \sum_{j=1}^{i-1} e_j=0\right\} \bmod (w+1)
\end{eqnarray*}
is the distance from the current time instant to the earliest energy arrival in the lookahead window or zero if there is no energy arrival in the window.
Based on $D_\tau$, the design of optimal $\mathcal{A}$ can be divided into two cases:

1) Nonzero energy arrivals observed in the lookahead window.
In this case, the TX knows that the battery will get fully charged at time $\tau+D_\tau$, so the planning of the energy usage is more like the offline mode.
It is clear that $1\le D_\tau\le w$.
Since the reward function is concave, the optimal policy is simply uniformly allocating the battery energy $B_\tau$ over the time slots from $\tau$ to $\tau+D_\tau-1$.

2) No energy arrival in the lookahead window.
In this case, $D_\tau=0$, indicating that the battery will not be fully charged until sometime  after $\tau+w$, and consequently the TX can only make decisions according to the current battery state $B_\tau$, working in a manner more like the online mode.

In summary, an optimal stationary policy must have the following form:
\begin{eqnarray}
\multicolumn{3}{l}{\mathcal{A}(b_\tau,e_{\tau+1},\ldots,e_{\tau+w})} \nonumber\\
\quad &= &\begin{cases}
\displaystyle\frac{b_\tau}{d_\tau}, &if $d_\tau\eqdef\mathcal{D}((e_{\tau+i})_{i=1}^w)>0$,\\[0.75ex]
\mathcal{A}(b_\tau,0,\ldots,0), &otherwise.
\end{cases}\label{eq:optimal-policy-Bern}
\end{eqnarray}

Suppose that the battery is fully charged at  time $\tau_1$.
Policy $\mathcal{A}$ will specify the energy expenditure $\xi_i^*$ in time slot $\tau_1+i-1$ for every $i\ge 1$ until time $\tau_2>\tau_1$ when an energy arrival is observed in the lookahead window, where
\begin{equation}\label{def:xi-Sequence}
  \xi^*_i \triangleq \mathcal{A}(b_i, 0, \ldots, 0)\quad\text{for $i\in\mathds{N}$}
\end{equation}
with $b_1 = B$ and $b_{i+1} = b_i - \xi^*_i$.
It is clear that $\sum_{i=1}^\infty \xi^*_i\le B$.
Then, from time $\tau_1$ to $\tau_2+w-1$, we have the following two-stage energy allocation sequence:
\begin{eqnarray}[c]
A_{\tau_1}=\xi^*_1, \ldots, A_{\tau_2-1}=\xi^*_{\tau_2-\tau_1},\nonumber\\
A_{\tau_2}=\frac{b_{\tau_2}}{w}, \ldots, A_{\tau_2+w-1}=\frac{b_{\tau_2}}{w},\label{eq:allocationCycle}
\end{eqnarray}
where $b_{\tau_2}=B-\sum_{i=1}^{\tau_2-\tau_1} \xi^*_i$.
Note that the battery is fully charged again at time $\tau_2+w$, and hence a new cycle of energy allocation is started.
This routine is detailed in Algorithm~\ref{Fig:Algorithm}.
\begin{algorithm*}[t]
  \caption{Optimal stationary policy $\pi_s^*(w)$ for Bernoulli energy arrivals~\eqref{eq:EH-Seq-distribtion}}\label{optimal_policy_algorithm}
  \label{Fig:Algorithm}
  \begin{algorithmic}
 \Require{Window size $w$, battery capacity $B$, observed energy sequence $(E_t)_{t=1}^{\tau + w}$ at any time $\tau$, and sequence $(\xi_j^*)_{j=1}^\infty$.}
 \Ensure{The optimal assigned energy (action) $a^*_\tau$ at time $\tau$}.
  \textit{Initialize}: \State{Set time $\tau \leftarrow 1$,  distance $d \leftarrow 0$, battery level $b \leftarrow B$, counter $i\leftarrow 1$, and counter  $j\leftarrow 1$}.
  \Loop   \Comment{No observation $(d=0)$ by default}
  \While {($d = 0$ \AND $i \leq w$)}  
  \If {($e_{i+\tau} = B$)} \Comment{Find the distance to the next arrival}
  \State $d \leftarrow i$; $j\leftarrow 1$; $i\leftarrow 1$
  \EndIf
  \State $i \leftarrow i+1$
  \EndWhile
  \If {($d \ne 0$)} \Comment{If an arrival is observed}
  \State $a_\tau \leftarrow \frac{b}{d}$; $d \leftarrow d-1$
  \Else \Comment{If no arrival is observed}
  \State $a_\tau \leftarrow \xi^*_j$; $j \leftarrow j+1$; $i \leftarrow w$
  \EndIf
  \State $b \leftarrow \min\{b - a_\tau + e_{\tau+1}, B\}$ \Comment{Battery level is updated}
  \State $\tau \leftarrow \tau + 1$
  \EndLoop
 \end{algorithmic}
\end{algorithm*}
The complexity of this algorithm is $\mathcal{O}(N)$, where $N$ is the length of energy arrival sequence.
Now, the design of optimal $\mathcal{A}$ boils down to determining the optimal sequence $(\xi_i^*)_{i=1}^\infty$, which is the main result of this paper.

\begin{theorem}%
[Theorems~\ref{Theorem:Main-Result} and \ref{Theorem:Main-Result2} and Propositions~\ref{pr:optimal-energy-general-equation} and \ref{pr:gap}]%
\label{th:summary}
The optimal sequence $(\xi^*_i)_{i=1}^\infty$ is the unique solution of
\[
\begin{cases}
\displaystyle \mathcal{R}'(x_i)
= p\mathcal{R}'\left(\frac{B-\sum_{j=1}^i x_j}{w}\right) + (1-p)\mathcal{R}'(x_{i+1}),\\
\multicolumn{1}{r}{i\in\pinteger,}\\
\displaystyle \sum_{i=1}^\infty x_i
= B.
\end{cases}
\]
The corresponding maximum average throughput $\Gamma^*=\mathcal{T}_\infty((\xi^*_i)_{i=1}^\infty)$ is given by \eqref{eq:opt.throughput}.
The finite sequences $(\underline{\xi}^{(N)*}_i)_{i=1}^N$ and $(\overline{\xi}^{(N)*}_i)_{i=1}^N$ (as the unique solutions of \eqref{eq:lower.opt.kkt} and \eqref{eq:upper.opt.kkt}, respectively) provide an asymptotic approximation of $(\xi^*_i)_{i=1}^N$ from above and below, respectively.
That is,
\[
\overline{\xi}^{(N)*}_i
< \xi^*_i
< \underline{\xi}^{(N)*}_i
\quad\text{for $N\ge i$}
\]
and
\[
\lim_{N\to\infty} \underline{\xi}^{(N)*}_i
= \lim_{N\to\infty} \overline{\xi}^{(N)*}_i
= \xi^*_i.
\]
Moreover, they yield a lower bound $\underline{\mathcal{T}}_N((\underline{\xi}^{(N)*}_i)_{i=1}^N)$ (Eq.~\eqref{eq:lower.opt.def}) and an upper bound $\overline{\mathcal{T}}_N((\overline{\xi}^{(N)*}_i)_{i=1}^N)$ (Eq.~\eqref{eq:upper.opt.def}) on the maximum average throughput, respectively, with the gap between these two bounds converging to zero as $N\to\infty$.
\end{theorem}

We end this section with a characterization of the maximum average throughput~\eqref{eq:long-term-throughput} in terms of $(\xi_i^*)_{i=1}^\infty$.
Let $F_0=1$ and $F_n$ (for $n\ge 1$) be the instant at which the $n$th energy $B$ (starting from $\tau=2$) arrives, that is,
\begin{equation}
F_n
\eqdef \min\left\{\tau'\ge 1:\sum_{\tau=2}^{\tau'} 1\{E_\tau=B\}\ge n\right\}.\label{fully_charged}
\end{equation}
It is clear that $(F_n)_{n=0}^\infty$ is a pure (or zero-delayed) renewal process.
At each instant $F_n$, the battery gets fully charged (either because the initial energy level $\beta=B$ for $n=0$ or because the arrival of energy $E_{F_n}=B$ for $n\ge 1$), and the post-$F_n$ process
\[
((F_{i+1}-F_i)_{i\ge n},(S_\tau)_{\tau\ge F_n})
\]
is independent of $(F_0,\ldots,F_n)$ and its distribution does not depend on $n$ as long as the policy is Markovian and time-invariant.
Then by definition (\cite[p.~169]{asmussen_applied_2003}), $(S_\tau)_{\tau=1}^\infty$ is a pure regenerative process with $(F_n)_{n=0}^\infty$ as regeneration points.
This observation naturally leads to the following result.

\begin{prop}\label{pr:optimal-energy-general-equation}
The maximum average throughput $\Gamma^*$ is given by
\begin{eqnarray}
\Gamma^*
&= &\mathcal{T}_\infty((\xi_i^*)_{i=1}^\infty) \nonumber\\
&\eqdef &\sum\limits_{k=1}^w p^2(1-p)^{k-1} k\mathcal{R}\left(\frac{B}{k}\right) \nonumber\\
& &+\sum_{k=1}^\infty p(1-p)^{k+w-1} \mathcal{R}(\xi^*_k) \nonumber\\
& &+\sum_{k=1}^\infty p^2(1-p)^{k+w-1} w \mathcal{R}\left(\frac{B-\sum_{j=1}^{k} \xi^*_j}{w}\right).
\label{eq:opt.throughput}
\end{eqnarray}
\end{prop}

\section{A Finite-Dimensional Approximation Approach for Characterizing the optimal $(\xi^*_i)_{i=1}^\infty$}\label{sec:proof}

Now we proceed to characterize the optimal sequence $(\xi_i^*)_{i=1}^\infty$, which is needed for the implementation of Algorithm~\ref{optimal_policy_algorithm} and the computation of the maximum average throughput $\Gamma^*$.

It is easy to see that any sequence $(\xi_i)_{i=1}^\infty$ satisfying
\begin{align*}
&\xi_i\ge 0, \quad i=1,2,\ldots,\nonumber\\
&\sum_{i=1}^\infty \xi_i \le B
\end{align*}
is associated with a valid stationary policy, and the induced average throughput is given by $\mathcal{T}_\infty((\xi_i)_{i=1}^\infty)$.
In view of Proposition~\ref{pr:optimal-energy-general-equation}, the optimal sequence $(\xi_i^*)_{i=1}^\infty$ must be the (unique) maximizer of the following convex optimization problem:
\begin{eqnarray}[Tlql]
maximize &\mathcal{T}_\infty((\xi_i)_{i=1}^\infty) \label{eq:opt}\\
subject to &\xi_i\ge 0, \quad i=1,2,\ldots,\nonumber\\
&\sum_{i=1}^\infty \xi_i \le B.\nonumber
\end{eqnarray}
However, (\ref{eq:opt}) is an infinite-dimensional optimization problem, which is not amenable to direct analysis.
We shall overcome this difficulty by constructing lower and upper bounds on \eqref{eq:opt} that involve only the $N$-length truncation $(\xi_i)_{i=1}^N$ and are asymptotically tight as  $N\to\infty$. It will be seen that the constructed bounds have clear operational meanings and the relevant analyses shed considerable light on the properties of the optimal sequence $(\xi_i^*)_{i=1}^\infty$. Moreover, as a byproduct, our work yields an efficient numerical method to approximate both the maximum average throughput and the optimal policy with guaranteed accuracy. 

\subsection{A Lower Bound}\label{sec:lower}

We construct a lower bound by adding to \eqref{eq:opt} an extra constraint $\xi_i=0$ for $i>N\in\pinteger$.
Intuitively, this constraint prohibits consuming energy after the first $N$ time slots in a renewal cycle until the next energy arrival is observed in the lookahead window. In this way, (\ref{eq:opt}) is reduced to a finite-dimensional convex optimization problem:
\begin{eqnarray}[Tlql]
maximize &\underline{\mathcal{T}}_N((\xi_i)_{i=1}^N) \label{eq:lower.opt}\\
subject to &\xi_i\ge 0, \quad i=1,2,\ldots,N,\nonumber\\
&\sum_{i=1}^N \xi_i \le B,\nonumber
\end{eqnarray}
where
\begin{eqnarray}
\multicolumn{3}{l}{\underline{\mathcal{T}}_N((\xi_i)_{i=1}^N)}\nonumber\\
\hspace{0.8em} &\eqdef &\mathcal{T}_\infty(\xi_1,\ldots,\xi_N,0,\ldots) \nonumber\\
&= &\sum\limits_{k=1}^w p^2(1-p)^{k-1} k\mathcal{R}\left(\frac{B}{k}\right) + \sum_{k=1}^N p(1-p)^{k+w-1} \mathcal{R}(\xi_k) \nonumber\\
& &+\sum_{k=1}^{N-1} p^2(1-p)^{k+w-1} w \mathcal{R}\left(\frac{B-\sum_{j=1}^{k} \xi_j}{w}\right) \nonumber\\
& &+p(1-p)^{N+w-1}w \mathcal{R}\left(\frac{B-\sum_{j=1}^{N} \xi_j}{w}\right).\label{eq:lower.opt.def}
\end{eqnarray}
Clearly, \eqref{eq:lower.opt} provides a lower bound on the maximum average throughput.
Moreover, the optimal sequence $(\xi_i^*)_{i=1}^\infty$ can be determined by first solving \eqref{eq:lower.opt} and then sending $N\to\infty$.

\begin{theorem}\label{th:lower.opt.kkt}
The (unique) maximizer $(\underline{\xi}^{(N)*}_i)_{i=1}^N$ of \eqref{eq:lower.opt} is the unique solution of
\begin{subequations}\label{eq:lower.opt.kkt}
\begin{eqnarray}
\mathcal{R}'(\underline{\xi}^{(N)*}_i)
&= &p\mathcal{R}'\left(\frac{B-\sum_{j=1}^i \underline{\xi}^{(N)*}_j}{w}\right) + (1-p)\mathcal{R}'(\underline{\xi}^{(N)*}_{i+1}),\nonumber\\
& &\multicolumn{1}{r}{1\le i<N,\hspace{3em}} \label{eq:lower.opt.kkt.a}\\
\underline{\xi}^{(N)*}_N
&= &\frac{B-\sum_{j=1}^N \underline{\xi}^{(N)*}_j}{w}.\label{eq:lower.opt.kkt.b}
\end{eqnarray}
\end{subequations}
Furthermore, $\underline{\xi}^{(N)*}_i$ is positive, strictly decreasing in $i$, and  satisfies
\[
\underline{\xi}^{(N)*}_i
< \frac{B-\sum_{j=1}^i \underline{\xi}^{(N)*}_j}{w} \quad\text{for $1\le i< N$}.
\]
\end{theorem}

To facilitate the asymptotic analysis of $(\underline{\xi}^{(N)*}_i)_{i=1}^N$ as $N\to\infty$, we define $\underline{\xi}^{(N)*}_i\eqdef 0$ if $i>N$; as a consequence, we have $(\underline{\xi}^{(N)*}_i)_{i=1}^\infty=(\underline{\xi}^{(N)*}_1,\ldots,\underline{\xi}^{(N)*}_N,0,0,\ldots)$.

\begin{theorem}\label{Theorem:Main-Result}
The sequence $(\underline{\xi}^{(j)*}_i)_{j=1}^\infty$ is strictly decreasing in $j$ for $j\ge i$, and
\[
\underline{\xi}^{*}_i
\eqdef \lim_{j\to\infty} \underline{\xi}^{(j)*}_i
\]
exists.
Moreover, $(\underline{\xi}^*_i)_{i=1}^\infty$ is the unique solution of
\begin{subequations}\label{eq:opt.kkt}
\begin{eqnarray}
\mathcal{R}'(x_i)
&= &p\mathcal{R}'\left(\frac{B-\sum_{j=1}^i x_j}{w}\right) + (1-p)\mathcal{R}'(x_{i+1}),\nonumber\\
& &\multicolumn{1}{r}{i\in\pinteger,\quad}\label{eq:opt.kkt.a}\\
\sum_{i=1}^\infty x_i
&= &B,\label{eq:opt.kkt.b}
\end{eqnarray}
\end{subequations}
and $\underline{\xi}^*_i=\xi^*_i$ for all $i\in\pinteger$ (that is, $(\underline{\xi}^*_i)_{i=1}^\infty$ is the unique maximizer of \eqref{eq:opt}), where $(x_i)_{i=1}^\infty$ denotes a sequence of nonnegative real numbers.
In addition, $\xi^*_i$ is positive, strictly decreasing in $i$, and  satisfies
\[
\xi^*_i
< \frac{B-\sum_{j=1}^i \xi^*_j}{w}.
\]
\end{theorem}

\begin{remark}
	As energy consumption is not permitted 
	after the first $N$ time slots in a renewal cycle until the next energy arrival is observed in the lookahead window,
	the TX is inclined to expend energy more aggressively in those permissible time slots. 
	This provides an intuitive explanation why $\underline{\xi}^{(N)*}_i$ converges from above to $\xi^*_i$ as $N\rightarrow\infty$.
\end{remark}

\begin{remark}
As $w\to\infty$, the maximum average throughput of the studied model converges to that of the offline model, which is given by
\begin{equation*}
\left.\Gamma^*\right|_{w=\infty}
= \sum_{k=1}^{\infty} p^2(1-p)^{k-1}k\mathcal{R}\left(\frac{B}{k}\right).
\end{equation*}
On the other hand, if $w=0$, $\Gamma^*$ reduces to the maximum average throughput of the online model:
\[
\left.\Gamma^*\right|_{w=0}
= \sum_{k=1}^\infty p(1-p)^{k-1} \mathcal{R}(\xi^*_k)
\quad\text{(\cite[Appx.~C]{ozgur16})}.
\]
The optimal $(\xi^*_i)_{i=1}^\infty$ for $w=0$ is however determined by
\begin{subequations}
\begin{eqnarray}
\mathcal{R}'(x_i)
&= &(1-p)\mathcal{R}'(x_{i+1}),
\quad 1\le i< M,\\
\mathcal{R}'(x_M)
&\ge &(1-p)\mathcal{R}'(0),\\
\sum_{i=1}^M x_i
&= &B,
\end{eqnarray}
\end{subequations}
instead of \eqref{eq:opt.kkt}, where $M$ is also implicitly determined by the equations.
Note that there is a drastic difference between the optimal $(\xi^*_i)_{i=1}^\infty$ for $w=0$ and the optimal $(\xi^*_i)_{i=1}^\infty$ for $w>0$: the latter is a strictly positive sequence while the former is positive only for the first $M$ entries.
This difference admits the following intuitive explanation. For online power control, the TX should deplete the battery once its energy level is below a certain threshold because the potential loss caused by battery overflow (as the saved energy may get wasted if the battery is fully charged in the next time slot) outweighs 
the benefit of keeping a small amount of energy in the battery. In contrast, with the ability to lookahead, the TX can handle the battery overflow issue more effectively by switching to the uniform allocation scheme once the next energy arrival is seen in the lookahead window and is under no pressure to deplete the battery before that.
\end{remark}

\subsection{An Upper Bound}\label{sec:upper}

We have characterized the optimal sequence $(\xi^*_i)_{i=1}^\infty$ by finding a finite sequence $(\underline{\xi}^{(N)*}_i)_{i=1}^N$ converging from above to $(\xi^*_i)_{i=1}^\infty$. Note that $(\underline{\xi}^{(N)*}_i)_{i=1}^N$ can be efficiently computed by applying convex optimization solvers to \eqref{eq:lower.opt}. 
However, due to a lack of knowledge of convergence rate, it is unclear how close $(\underline{\xi}^{(N)*}_i)_{i=1}^N$ is to $(\xi^*_i)_{i=1}^\infty$ for a given $N$. As a remedy, we shall construct a finite sequence converging from below to $(\xi^*_i)_{i=1}^\infty$, which provides an alternative way to approximate $(\xi^*_i)_{i=1}^\infty$ and, more importantly, enables us to reliably estimate the precision of the approximation results.

Note that $(\underline{\xi}^{(N)*}_i)_{i=1}^N$ (which converges from above to $(\xi^*_i)_{i=1}^\infty$) is obtained by analyzing a lower bound on the maximum average throughput. This suggests that
in order to get a finite sequence converging from below to $(\xi^*_i)_{i=1}^\infty$, we may need  to construct a suitable upper bound on the maximum average throughput. 
To this end, we first express $\mathcal{T}_\infty((\xi_i)_{i=1}^\infty)$
in the following equivalent form:
\begin{eqnarray}
\mathcal{T}_\infty((\xi_i)_{i=1}^\infty)&=&  \sum_{k=1}^w p^2(1-p)^{k-1} k \mathcal{R}\left(\frac{B}{k}\right) \nonumber\\
& &+ \sum_{k=w+1}^\infty p^2(1-p)^{k-1} \Bigg( \sum_{j=1}^{k-w} \mathcal{R}(\xi_j) \nonumber\\
& &+ w \mathcal{R}\left(\frac{B-\sum_{j=1}^{k-w} \xi_j}{w}\right) \Bigg). \label{eq:equivalentT}
\end{eqnarray}
By Jensen's inequality, it is easy to show that, for $k>N+w$,
\begin{equation}
\sum_{j=1}^{k-w} \mathcal{R}(\xi_j) + w \mathcal{R}\left(\frac{B-\sum_{j=1}^{k-w} \xi_j}{w}\right) \label{eq:replaced}
\end{equation}
can be bounded above by
\begin{equation}
\sum_{j=1}^N \mathcal{R}(\xi_j) + (k-N) \mathcal{R}\left(\frac{B-\sum_{j=1}^{N} \xi_j}{k-N}\right)\label{eq:critical.upperbound}.
\end{equation}
One can gain an intuitive understanding of~\eqref{eq:critical.upperbound} by considering the following scenario: within each renewal cycle, if after the first $N$ time slots, there is still no energy arrival  observed in the lookahead window, then a genie will inform the TX  the next arrival time so that it can adopt the optimal uniform allocation scheme. It should be clear that~\eqref{eq:critical.upperbound} is exactly the genie-aided throughput in a renewal cycle of length $k$. Substituting~\eqref{eq:replaced} with~\eqref{eq:critical.upperbound} in~\eqref{eq:equivalentT} 
yields, after some algebraic manipulations, the following upper bound on  $\mathcal{T}_\infty((\xi_i)_{i=1}^\infty)$:
\begin{eqnarray}
\multicolumn{3}{l}{\mathcal{T}_\infty((\xi_i)_{i=1}^\infty) \leq \overline{\mathcal{T}}_N((\xi_i)_{i=1}^N)} \nonumber\\
\quad &\eqdef &\sum\limits_{k=1}^w p^2(1-p)^{k-1} k\mathcal{R}\left(\frac{B}{k}\right) \nonumber\\
& &+\sum_{k=1}^N p(1-p)^{k+w-1} \mathcal{R}(\xi_k) \nonumber\\
& &+\sum_{k=1}^{N-1} p^2(1-p)^{k+w-1} w \mathcal{R}\left(\frac{B-\sum_{j=1}^{k} \xi_j}{w}\right) \nonumber\\
& &+\sum_{k=w}^\infty p^2(1-p)^{k+N-1} k \mathcal{R}\left(\frac{B-\sum_{j=1}^{N} \xi_j}{k}\right). \nonumber\\
\label{eq:upper.opt.def}
\end{eqnarray}
Replacing $\mathcal{T}_\infty((\xi_i)_{i=1}^\infty)$ with $\overline{\mathcal{T}}_N((\xi_i)_{i=1}^N)$ in~\eqref{eq:opt}
gives a finite-dimensional convex optimization problem (which provides an upper bound on the maximum average throughput):
\begin{eqnarray}[Tlql]
maximize &\overline{\mathcal{T}}_N((\xi_i)_{i=1}^N) \label{eq:upper.opt}\\
subject to &\xi_i\ge 0, \quad i=1,2,\ldots,N,\nonumber\\
&\sum_{i=1}^N \xi_i \le B.\nonumber
\end{eqnarray}

\begin{theorem}\label{th:upper.opt.kkt}
The (unique) maximizer $(\overline{\xi}^{(N)*}_i)_{i=1}^N$ of \eqref{eq:upper.opt} is the unique solution of
\begin{subequations}\label{eq:upper.opt.kkt}
\begin{eqnarray}[rcl]
\mathcal{R}'(\overline{\xi}^{(N)*}_i)
&= &p\mathcal{R}'\left(\frac{B-\sum_{j=1}^i \overline{\xi}^{(N)*}_j}{w}\right) + (1-p)\mathcal{R}'(\overline{\xi}^{(N)*}_{i+1}), \nonumber\\
& &\multicolumn{1}{r}{1\le i<N, \hspace{3em}} \label{eq:upper.opt.kkt.a}\\
\mathcal{R}'(\overline{\xi}^{(N)*}_N)
&= &\sum_{k=w}^\infty p(1-p)^{k-w} \mathcal{R}'\left(\frac{B-\sum_{j=1}^N \overline{\xi}^{(N)*}_j}{k}\right). \nonumber\\
\label{eq:upper.opt.kkt.b}
\end{eqnarray}
\end{subequations}
Furthermore, $\overline{\xi}^{(N)*}_i$ is positive, strictly decreasing in $i$, and satisfies
\[
\overline{\xi}^{(N)*}_i
< \frac{B-\sum_{j=1}^i \overline{\xi}^{(N)*}_j}{w} \quad\text{for $1\le i\le N$}.
\]
\end{theorem}

To facilitate the asymptotic analysis of $(\overline{\xi}^{(N)*}_i)_{i=1}^N$ as $N\to\infty$, we define $\overline{\xi}^{(N)*}_i\eqdef 0$ if $i>N$; as a consequence, we have $(\overline{\xi}^{(N)*}_i)_{i=1}^\infty=(\overline{\xi}^{(N)*}_1,\ldots,\overline{\xi}^{(N)*}_N,0,0,\ldots)$.

\begin{theorem}\label{Theorem:Main-Result2}
The sequence $(\overline{\xi}^{(j)*}_i)_{j=1}^\infty$ is strictly increasing in $j$ for $j\ge i$, and
\[
\overline{\xi}^{*}_i
\eqdef \lim_{j\to\infty} \overline{\xi}^{(j)*}_i
\]
exists.
Moreover, $(\overline{\xi}^*_i)_{i=1}^\infty$ is the unique solution of \eqref{eq:opt.kkt} and hence $\overline{\xi}^*_i=\xi^*_i$ for all $i\in\pinteger$.
\end{theorem}

\begin{remark}
	With the availability of information from the genie, 	
	the TX will be able to expend energy more efficiently by  switching to the optimal uniform allocation scheme; as a consequence, it is inclined to be more conservative in the first $N$ time slots within each renewal cycle when there is  no energy arrival  observed in the lookahead window. This provides an intuitive explanation why $\overline{\xi}^{(N)*}_i$ converges from below to $\xi^*_i$ as $N\rightarrow\infty$.
\end{remark}

Theorem~\ref{Theorem:Main-Result2} establishes a useful upper bound on the maximum average throughput as well as a lower bound on the optimal sequence $(\xi^*_i)_{i=1}^\infty$.
The next result investigates the gap between the upper bound $\overline{\Gamma}_N^*\eqdef \overline{\mathcal{T}}_N((\overline{\xi}^{(N)*}_i)_{i=1}^N)$ and the lower bound $\underline{\Gamma}_N^*\eqdef \underline{\mathcal{T}}_N((\underline{\xi}^{(N)*}_i)_{i=1}^N)$.
It shows that the gap converges to zero as $N\to\infty$ or $w\to\infty$

\begin{prop}\label{pr:gap}
\[
\overline{\Gamma}_N^* - \underline{\Gamma}_N^*
< p(1-p)^{N+w}\mathcal{R}'(0)\left(B-\sum_{j=1}^N \overline{\xi}^{(N)*}_j\right).
\]
\end{prop}

\begin{remark}
	For the online model (i.e., $w=0$), with $\overline{\mathcal{T}}_N((\xi_i)_{i=1}^N)$ and $\underline{\mathcal{T}}_N((\xi_i)_{i=1}^N)$ taking the following degenerate forms
	\begin{align*}
	\overline{\mathcal{T}}_N((\xi_i)_{i=1}^N)&\eqdef\sum\limits_{k=1}^Np(1-p)^{k-1}\mathcal{R}(\xi_k)\\
	&\quad +\sum\limits_{k=1}^{\infty}p^2(1-p)^{k+N-1}k\mathcal{R}\left(\frac{B-\sum_{j=1}^N\xi_j}{k}\right),\\
	\underline{\mathcal{T}}_N((\xi_i)_{i=1}^N)&\eqdef\sum\limits_{k=1}^Np(1-p)^{k-1}\mathcal{R}(\xi_k),
	\end{align*}
	the relationship $\underline{\Gamma}_N^*\leq\Gamma^*\leq\overline{\Gamma}_N^*$ remains valid. Moreover, it is easy to verify that Proposition \ref{pr:gap}  continues to hold when $w=0$.
\end{remark}

\section{Numerical Results}\label{sec:numerical}

Here we present some numerical results to illustrate our theoretical findings.

\begin{figure}[htbp]
	\centering
	\includegraphics{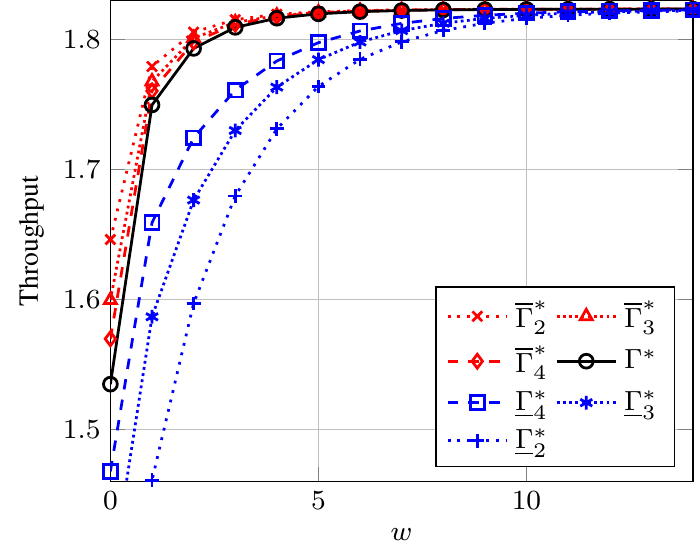}
	\caption{The maximum average throughput $\Gamma^*$, the upper bound $\overline{\Gamma}_N^*$, and  the lower bound $\underline{\Gamma}_N^*$
	versus the lookahead window size $w$ for $B=100$, $p=0.3$, and $\gamma=0.5$.}\label{fig:Throughput-window}
\end{figure}

In Fig.~\ref{fig:Throughput-window}, we 
plot the maximum average throughput $\Gamma^*$ against the lookahead window size $w$ for the given system parameters. Clearly, $\Gamma^*$ is a monotonically increasing function of $w$
with $\left.\Gamma^*\right|_{w=0}$ being the maximum average throughput of the online model and $\left.\Gamma^*\right|_{w=\infty}$ being the maximum average throughput of the offline model, respectively. Note that it suffices to have the lookahead window size $w=5$ to achieve over $99.5\%$ of the maximum average throughput of the offline model.
For comparisons, we also plot the upper bound $\overline{\Gamma}_N^*$ and the lower bound $\underline{\Gamma}_N^*$. It can be seen that both bounds converge to $\Gamma^*$ as $N\rightarrow\infty$ (with $w$ fixed) and converge to $\left.\Gamma^*\right|_{w=\infty}$ as $w\rightarrow\infty$ (with $N$ fixed), which is consistent with 
Proposition \ref{pr:gap}.

\begin{figure}[ht!]
	\centering
	\includegraphics{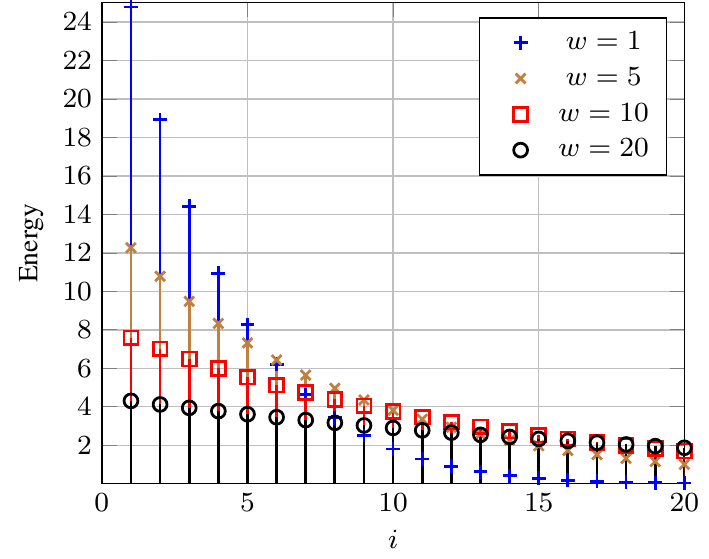}
	\caption{The optimal sequences  $(\xi^*_i)_{i=1}^{\infty}$
	associated with different lookahead window sizes $w$ for $B=100$, $p=0.3$, and $\gamma=0.5$.
		}\label{fig: optimal-Seq}
\end{figure}

Fig. \ref{fig: optimal-Seq} illustrates 
 the optimal sequences  $(\xi^*_i)_{i=1}^{\infty}$ associated with different lookahead window sizes $w$ for the given system parameters. It can be seen that for a fixed $w$, $\xi^*_i$ is  strictly decreasing in $i$, as indicated by Theorem \ref{Theorem:Main-Result}. On the other hand, for a fixed $i$, $\xi^*_i$ is not necessarily monotonic with respect to $w$. In general, as $w$ increases, the optimal sequence $(\xi^*_i)_{i=1}^{\infty}$ gradually shift its weight towards the right.

\begin{figure*}[ht!]
	\centering
	\includegraphics{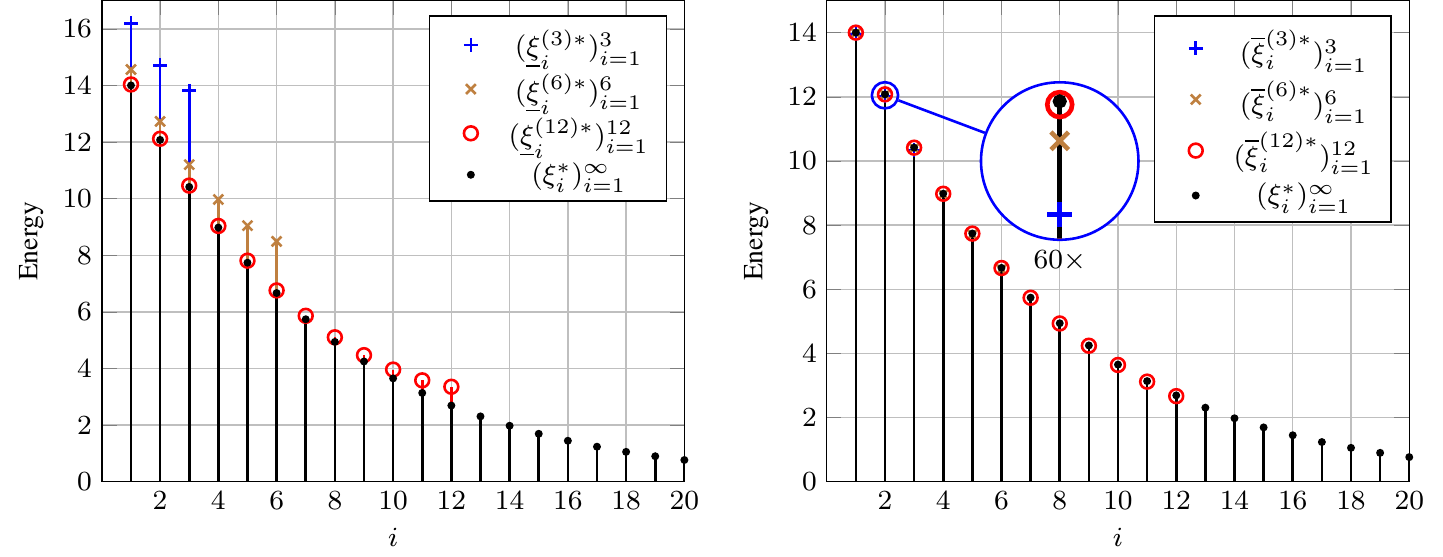}
	\caption{Comparisons of the optimal sequence $(\xi^*_i)_{i=1}^{\infty}$ with $(\underline{\xi}_i^{(N)*})_{i=1}^N$ and $(\overline{\xi}_i^{(N)*})_{i=1}^N$  for $B=100$, $p=0.3$, $\gamma=0.5$, and $w=4$.}\label{fig: Seq-bounds}
\end{figure*}

In Fig.~\ref{fig: Seq-bounds}, we compare the optimal sequence $(\xi^*_i)_{i=1}^{\infty}$ with various $(\underline{\xi}_i^{(N)*})_{i=1}^N$ and
$(\overline{\xi}_i^{(N)*})_{i=1}^N$ for the given system parameters. As expected from Theorems \ref{Theorem:Main-Result} and \ref{Theorem:Main-Result2}, $\underline{\xi}_i^{(N)*}$ and $\overline{\xi}_i^{(N)*}$ converge to $\xi^*_i$ from above and below, respectively, as $N\rightarrow\infty$. One surprising phenomenon is that $\overline{\xi}_i^{(N)*}$ (with $N\geq i$) is almost indistinguishable from $\xi^*_i$ even when $N$ is very small.
This is possibly because the extra information provided by the genie has a much milder impact on the TX's optimal energy consumption behaviour 
as compared to setting a forbidden time interval, rendering $\overline{\xi}_i^{(N)*}$ a better approximation of $\xi^*_i$ than $\underline{\xi}_i^{(N)*}$ for the same $N$.

The significance of the optimal sequence $(\xi^*_i)_{i=1}^{\infty}$ is not restricted to the case of Bernoulli energy arrivals because effective policies for a more general case can be designed based on it.
The Bernoulli-optimal policy \eqref{eq:optimal-policy-Bern} is essentially a combination of the optimal offline policy (see, e.g., \cite{ulukus,TY12}) and an ``optimal online'' policy for the case of nonzero energy arrivals and the case of zero energy arrivals in the lookahead window, respectively.
It is then natural to imagine an extension of \eqref{eq:optimal-policy-Bern}, still a combination of the optimal offline policy and an ``online'' policy, one for the case of ``enough'' energy arrivals in the lookahead window and the other for the case of ``close-to-zero'' energy arrivals in the lookahead window.
The continuity of the battery updating law \eqref{eq:battery-change-relation} and the reward function \eqref{eq:reward-per-slot} ensures some degree of continuity of the problem in the vicinity of a Bernoulli distribution, so it is reasonable to expect a good performance of the ``online-policy'' part $\mathcal{A}(b_\tau,0,\ldots,0)$ of \eqref{eq:optimal-policy-Bern} in the case of close-to-zero energy arrivals in the lookahead window.
So far, however, $\mathcal{A}(b_\tau,0,\ldots,0)$ is uniquely defined only for some discrete battery energy levels, i.e., the optimal sequence $(\xi^*_i)_{i=1}^{\infty}$ defined by \eqref{def:xi-Sequence}, so one obstacle is how to extend the support of this policy to cover all possible battery energy levels.
This can be solved by considering the function relation between $\xi^*_1=\mathcal{A}(B,0,\ldots,0)$ and $B$, a method used in \cite{YC20}.

Inspired by the ideas above, we extend the Bernoulli-optimal policy \eqref{eq:optimal-policy-Bern} to a policy for general i.i.d.\ energy arrivals with a distribution $Q$.
It is defined as follows:
\begin{eqnarray}
\multicolumn{3}{l}{\omega_w(b_\tau,e_{\tau+1},\ldots,e_{\tau+w})} \nonumber\\
\hspace{0.5em} &\eqdef &\begin{cases}
\overline{\omega}_w(b_\tau,e_{\tau+1},\ldots,e_{\tau+w}),\\
\multicolumn{2}{r}{\text{if $\sum_{i=1}^w e_{\tau+i}\ge B/2$,}}\\
\min\{\underline{\omega}_w(b_\tau+\sum_{i=1}^w e_{\tau+i}),b_\tau\}, &otherwise.
\end{cases}
\label{eq:general-policy}
\end{eqnarray}
where $\overline{\omega}_w$ is the optimal offline policy with the knowledge of the current battery energy level and the energy arrivals in the lookahead window (of size $w$), and $\underline{\omega}_w(x)\eqdef\xi_1^*|_{B=x}$.
The associated parameter $p$ of $\underline{\omega}_w$ is set to be the mean-to-capacity ratio (MCR) of $Q$ (\cite[Def.~3]{YC20}), denoted as $\mcr(Q)$.
\begin{figure}[ht!]
	\centering
	\includegraphics{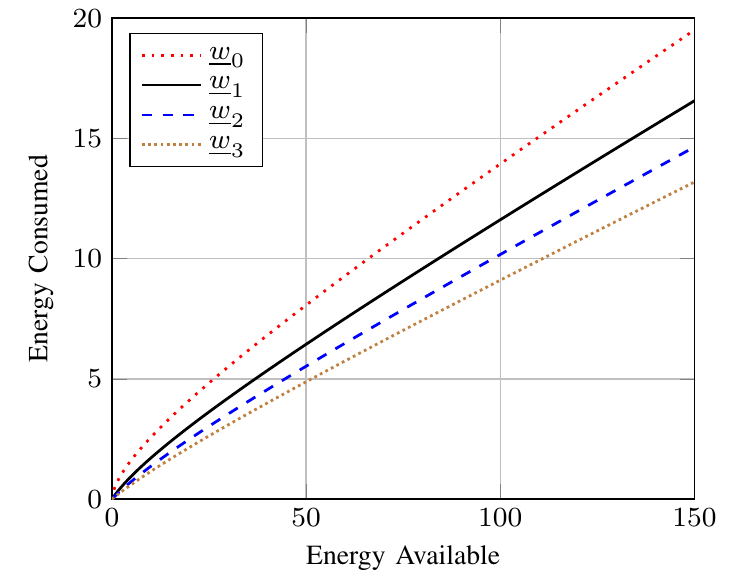}
	\caption{Plots of $\underline{\omega}_w$ for $w=0,1,2,3$ with $p=0.1$ and $\gamma=0.5$.}\label{fig:policy.ext}
\end{figure}
Fig.~\ref{fig:policy.ext} illustrates the graph of $\underline{\omega}_w$ for $w=0,1,2,3$ with $p=0.1$ and $\gamma=0.5$.
Policy $\underline{\omega}_0$ is just the maximin optimal policy in \cite{YC20}, a piecewise linear function.
\begin{figure}[ht!]
	\centering
	\includegraphics{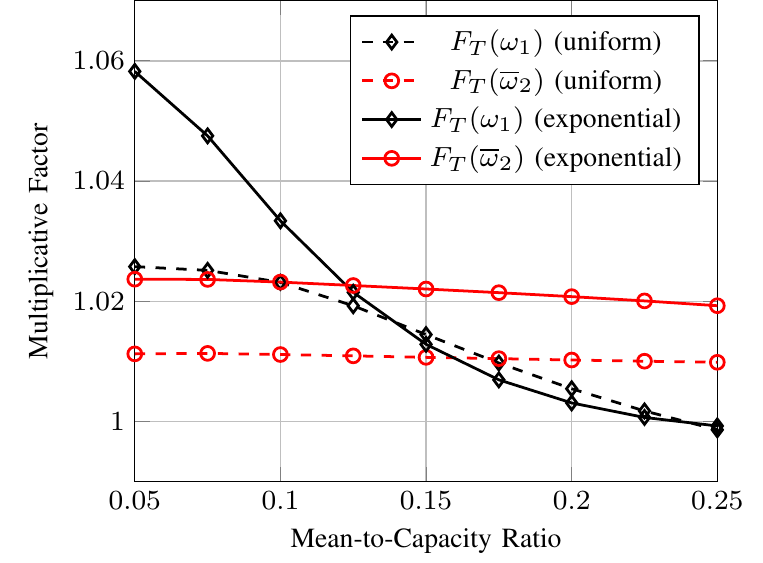}
	\caption{The multiplicative factors of $\omega_1$ and $\overline{\omega}_2$ for $B=100$, $\gamma=0.5$, $T=10^4$, and $Q$ uniform or exponential.}\label{fig:policy.perf}
\end{figure}
Fig.~\ref{fig:policy.perf} compares the performance of $\omega_1$, $\overline{\omega}_1$, and $\overline{\omega}_2$ when the energy arrival distribution $Q$ is uniform (over $[0,c]$ for some $c>0$) or exponential.
To facilitate the comparison, we choose the expected throughput induced by $\overline{\omega}_1$ as the baseline, and define the $T$-horizon multiplicative factor of a policy $\pi$ by
\[
F_T(\pi)
\eqdef \frac{\Gamma_T^{\pi}}{\Gamma_T^{\overline{\omega}_1}}.
\]
Note that $\omega_1$ outperforms $\overline{\omega}_1$ when $\mcr(Q)$ is approximately below 0.24, and even outperforms $\overline{\omega}_2$ for smaller MCRs.
The rationale behind this fact is that an optimal offline policy will use up all the energy at the end of the sequence of energy arrivals, so the policy $\overline{\omega}_w$ with the knowledge of energy arrivals in the lookahead window tends to be locally optimal, especially in the case of close-to-zero energy arrivals, although its action in the next time slot will change according to the updated lookahead window.

\section{Conclusion}  \label{Sec:Conclusion}

In this paper, we have introduced a new EH communication system model,
where the TX has the knowledge of future energy arrivals in a
lookahead window. This model provides a bridge between
the online model and the offline model, which have been
previously studied in isolation. A complete characterization of
the optimal power control policy is obtained for the new model
with Bernoulli energy arrivals, which is notably different from its counterpart for the online model. 

It is worth mentioning that the special reward function considered in this work only plays a marginal role in our analysis. Indeed,  
most of the results  in the present paper  require very weak conditions on the reward function. Therefore, they are not confined to the scenario where the objective is to maximize the throughput over an AWGN channel. We have also made some initial attempts to deal with general i.i.d. energy arrivals.
But obtaining an explicit characterization of the optimal power control policy beyond the Bernoulli case appears to be analytically difficult. Nevertheless, it can be shown that the maximum throughput under a generic energy arrival process can be closely approximated by that under a suitably constructed Bernoulli process in certain asymptotic regimes. More importantly, as suggested by the existing works on the online model \cite{ozgur16, ABU18, YC20}, due to the extremal property of the Bernoulli distribution, the relevant results have potentially very broad implications.
Investigating and exploring such implications, especially with respect to the design of reinforcement-learning-based schemes \cite{YA21} for more complex scenarios, is a worthwhile endeavor for future research.

\appendices

\section{Proofs of Results in Section~\ref{Sec:Strategy}}

\begin{IEEEproof}[Proof of Proposition~\ref{pr:optimal-energy-general-equation}]
Since $(S_\tau)_{\tau=1}^\infty$ is a pure regenerative process with $(F_n)_{n=0}^\infty$ (defined by \eqref{fully_charged}) as regeneration points, we define
\[
L_n = F_n-F_{n-1}
\]
for $n\ge 1$.
It is clear that $(L_n)_{n=1}^\infty$ is an i.i.d. sequence and
\[
P_{L_n}(k)
= P_{L}(k)
\eqdef p(1-p)^{k-1}.
\]
Then
\[
\expect L
= \sum_{k=1}^\infty kp(1-p)^{k-1}
= \frac{1}{p}
< +\infty
\]
and
\[
\expect\sum_{t=1}^L \mathcal{R}(A_t)
\le \mathcal{R}(B)\expect L
= \frac{\mathcal{R}(B)}{p}
< +\infty.
\]

By \cite[Lemma~1]{ozgur16},
\[
\lim_{T\to\infty} \frac{1}{T} \sum_{\tau=1}^T \mathcal{R}(A_\tau)
\to \frac{\expect\sum_{t=1}^L \mathcal{R}(A_t)}{\expect L}
\]
almost surely.
Also note that
\[
\frac{1}{T} \sum_{\tau=1}^T \mathcal{R}(A_\tau)
\le \mathcal{R}(B).
\]
By the dominated convergence theorem,
\begin{eqnarray}
\Gamma^*
&= &\expect \lim_{T\to\infty} \frac{1}{T} \sum_{\tau=1}^T \mathcal{R}(A_\tau)
= \frac{\expect\sum_{t=1}^L \mathcal{R}(A_t)}{\expect L} \nonumber\\
&= &p\sum_{k=1}^\infty p(1-p)^{k-1} \sum_{j=1}^k \mathcal{R}(A_j) \nonumber\\
&\eqvar{(a)} &p^2 \sum_{k=1}^w (1-p)^{k-1} k \mathcal{R}\left(\frac{B}{k}\right) \nonumber\\
& &+ p^2 \sum_{k=w+1}^\infty (1-p)^{k-1} \Bigg( \sum_{j=1}^{k-w} \mathcal{R}(\xi^*_j) \nonumber\\
& &+ w \mathcal{R}\left(\frac{B-\sum_{j=1}^{k-w} \xi^*_j}{w}\right) \Bigg) \nonumber\\
&= &p^2 \sum_{k=1}^w (1-p)^{k-1} k \mathcal{R}\left(\frac{B}{k}\right) \nonumber\\
& &+ p^2 \sum_{j=1}^\infty \mathcal{R}(\xi^*_j) \sum_{k=j+w}^\infty (1-p)^{k-1} \nonumber\\
& &+ p^2 \sum_{k=w+1}^\infty (1-p)^{k-1} w \mathcal{R}\left(\frac{B-\sum_{j=1}^{k-w} \xi^*_j}{w}\right) \nonumber\\
&= &p^2 \sum_{k=1}^w (1-p)^{k-1} k \mathcal{R}\left(\frac{B}{k}\right) \nonumber\\
& &+ p \sum_{j=1}^\infty (1-p)^{j+w-1} \mathcal{R}(\xi^*_j) \nonumber\\
& &+ p^2 \sum_{k=w+1}^\infty (1-p)^{k-1} w \mathcal{R}\left(\frac{B-\sum_{j=1}^{k-w} \xi^*_j}{w}\right), \nonumber
\end{eqnarray}
where (a) follows from \eqref{eq:allocationCycle}.
\end{IEEEproof}

\section{Proofs of Results in Section~\ref{sec:lower}}

\begin{IEEEproof}[Proof of Theorem~\ref{th:lower.opt.kkt}]
The uniqueness of the maximizer is an easy consequence of the strict concavity of $\mathcal{R}$.
The KKT conditions for $(\underline{\xi}^{(N)*}_i)_{i=1}^N$ are
\begin{eqnarray*}[l]
p(1-p)^{i+w-1}\mathcal{R}'(\underline{\xi}^{(N)*}_i)\\
\quad - \sum_{k=i}^{N-1} p^2(1-p)^{k+w-1} \mathcal{R}'\left(\frac{B-\sum_{j=1}^k \underline{\xi}^{(N)*}_j}{w}\right)\\
\quad - p(1-p)^{N+w-1} \mathcal{R}'\left(\frac{B-\sum_{j=1}^N \underline{\xi}^{(N)*}_j}{w}\right)\\
\quad + \underline{\mu}^{(N)}_i - \underline{\lambda}^{(N)}
= 0,\quad 1\le i\le N,\\
\underline{\mu}^{(N)}_i \underline{\xi}^{(N)*}_i
= 0,\quad 1\le i\le N,\\
\underline{\lambda}^{(N)}\left(B-\sum_{j=1}^N \underline{\xi}^{(N)*}_j\right)
= 0,
\end{eqnarray*}
where $\underline{\mu}^{(N)}_i$ and $\underline{\lambda}^{(N)}$ are all nonnegative.
Equivalently, we have
\begin{subequations}\label{lower.opt.kkt.2}
\begin{eqnarray}[l]
p(1-p)^{i+w-1} \mathcal{R}'(\underline{\xi}^{(N)*}_i) + \underline{\mu}_i^{(N)} \nonumber\\
\quad = p^2(1-p)^{i+w-1}\mathcal{R}'\left(\frac{B-\sum_{j=1}^i \underline{\xi}^{(N)*}_j}{w}\right) \nonumber\\
\qquad + p(1-p)^{i+w}\mathcal{R}'(\underline{\xi}^{(N)*}_{i+1}) + \underline{\mu}_{i+1}^{(N)}, \; 1\le i<N,\label{eq:lower.opt.kkt.2a}\\
p(1-p)^{N+w-1} \mathcal{R}'(\underline{\xi}^{(N)*}_N) + \underline{\mu}_N^{(N)} \nonumber\\
\quad = p(1-p)^{N+w-1} \mathcal{R}'\left(\frac{B-\sum_{j=1}^N \underline{\xi}^{(N)*}_j}{w}\right) + \underline{\lambda}^{(N)}, \nonumber\\
\label{eq:lower.opt.kkt.2b}\\
\underline{\mu}^{(N)}_i \underline{\xi}^{(N)*}_i
= 0,\quad 1\le i\le N,\label{eq:lower.opt.kkt.2c}\\
\underline{\lambda}^{(N)}\left(B-\sum_{j=1}^N \underline{\xi}^{(N)*}_j\right)
= 0\label{eq:lower.opt.kkt.2d}.
\end{eqnarray}
\end{subequations}

We first show that $\sum_{j=1}^N \underline{\xi}^{(N)*}_j < B$.
If $\sum_{j=1}^N \underline{\xi}^{(N)*}_j = B$, then
\begin{eqnarray*}[l]
p(1-p)^{N+w-1} \mathcal{R}'(\underline{\xi}^{(N)*}_N) + \underline{\mu}_N^{(N)}\\
\quad \ge p(1-p)^{N+w-1} \mathcal{R}'(0)
\quad\text{(Eq.~\eqref{eq:lower.opt.kkt.2b})},
\end{eqnarray*}
which implies $\underline{\xi}^{(N)*}_N=0$ because $\mathcal{R}'$ is strictly decreasing.
This further implies $\sum_{j=1}^{N-1} \underline{\xi}^{(N)*}_j = B$, so 
\begin{eqnarray*}[l]
p(1-p)^{N+w-2} \mathcal{R}'(\underline{\xi}^{(N)*}_{N-1}) + \underline{\mu}_{N-1}^{(N)}\\
\quad \ge p(1-p)^{N+w-2}\mathcal{R}'(0)
\quad\text{(Eq.~\eqref{eq:lower.opt.kkt.2a})},
\end{eqnarray*}
and consequently $\underline{\xi}^{(N)*}_{N-1}=0$.
Repeating this backward induction, we finally get $\underline{\xi}^{(N)*}_1=B=0$, which is absurd.
Therefore, $\sum_{j=1}^N \underline{\xi}^{(N)*}_j < B$. Now it can be readily seen  that $\underline{\lambda}^{(N)}=0$ (Eq.~\eqref{eq:lower.opt.kkt.2d}) and
\begin{eqnarray*}[l]
p(1-p)^{N+w-1} \mathcal{R}'(\underline{\xi}^{(N)*}_N) + \underline{\mu}_N^{(N)}\\
\quad < p(1-p)^{N+w-1} \mathcal{R}'(0)
\quad\text{(Eq.~\eqref{eq:lower.opt.kkt.2b})},
\end{eqnarray*}
which implies $\underline{\xi}^{(N)*}_N>0$ and $\underline{\mu}_N^{(N)}=0$ (Eq.~\eqref{eq:lower.opt.kkt.2c}).
So we have
\[
\mathcal{R}'(\underline{\xi}^{(N)*}_N)
= \mathcal{R}'\left(\frac{B-\sum_{j=1}^N \underline{\xi}^{(N)*}_j}{w}\right)
\quad\text{(Eq.~\eqref{eq:lower.opt.kkt.2b})},
\]
or \eqref{eq:lower.opt.kkt.b}.
Furthermore, since
\begin{eqnarray*}[l]
p(1-p)^{N+w-2} \mathcal{R}'(\underline{\xi}^{(N)*}_{N-1}) + \underline{\mu}_{N-1}^{(N)}\\
\quad < p(1-p)^{N+w-2}\mathcal{R}'(0)
\quad\text{(Eq.~\eqref{eq:lower.opt.kkt.2a})},
\end{eqnarray*}
it follows that $\underline{\xi}^{(N)*}_{N-1}>0$,  $\underline{\mu}_{N-1}^{(N)}=0$ (Eq.~\eqref{eq:lower.opt.kkt.2c}), and consequently
\[
\mathcal{R}'(\underline{\xi}^{(N)*}_{N-1})
= p\mathcal{R}'\left(\frac{B-\sum_{j=1}^{N-1} \underline{\xi}^{(N)*}_j}{w}\right) + (1-p)\mathcal{R}'(\underline{\xi}^{(N)*}_N).
\]
This further implies that
\begin{eqnarray*}
\mathcal{R}'(\underline{\xi}^{(N)*}_{N-1})
&< &p\mathcal{R}'\left(\frac{B-\sum_{j=1}^N \underline{\xi}^{(N)*}_j}{w}\right) + (1-p)\mathcal{R}'(\underline{\xi}^{(N)*}_N)\\
&\le &\mathcal{R}'(\underline{\xi}^{(N)*}_N)
\end{eqnarray*}
and
\begin{eqnarray*}
\mathcal{R}'(\underline{\xi}^{(N)*}_{N-1})
&\ge &p\mathcal{R}'\left(\frac{B-\sum_{j=1}^{N-1} \underline{\xi}^{(N)*}_j}{w}\right)\\
& &+ (1-p)\mathcal{R}'\left(\frac{B-\sum_{j=1}^N \underline{\xi}^{(N)*}_j}{w}\right)\\
&> &\mathcal{R}'\left(\frac{B-\sum_{j=1}^{N-1} \underline{\xi}^{(N)*}_j}{w}\right).
\end{eqnarray*}
Thus $\underline{\xi}^{(N)*}_{N-1}>\underline{\xi}^{(N)*}_N$ and
\[
\underline{\xi}^{(N)*}_{N-1}
< \frac{B-\sum_{j=1}^{N-1} \underline{\xi}^{(N)*}_j}{w}.
\]
Repeating such a backward induction finally establishes the theorem.
\end{IEEEproof}

\begin{IEEEproof}[Proof of Theorem~\ref{Theorem:Main-Result}]
From \eqref{eq:lower.opt.kkt.a}, it is observed that either
\[
\underline{\xi}^{(N)*}_i
> \underline{\xi}^{(N+1)*}_i \quad \text{for all $1\le i\le N$}
\]
or
\[
\underline{\xi}^{(N)*}_i
\le \underline{\xi}^{(N+1)*}_i \quad \text{for all $1\le i\le N$}.
\]
The latter is however impossible because we would have
\begin{eqnarray*}
\underline{\xi}^{(N+1)*}_N
&< &\frac{B-\sum_{j=1}^N \underline{\xi}^{(N+1)}_j}{w}
\le \frac{B-\sum_{j=1}^N \underline{\xi}^{(N)}_j}{w}
= \underline{\xi}^{(N)*}_N.
\end{eqnarray*}
This shows that $\underline{\xi}^{(j)*}_i$ is strictly decreasing in $j$ for $j\ge i$, which implies the existence of $\underline{\xi}^*_i$ and further gives \eqref{eq:opt.kkt.a} by the continuity of $\mathcal{R}'(x)$.
Since
\[
\underline{\xi}^*_i
< \underline{\xi}^{(i)*}_i
\le \frac{\sum_{j=1}^i \xi^{(i)*}_j}{i}
\le \frac{B}{i}
\to 0
\]
as $i\to\infty$, we must have $\sum_{i=1}^\infty \underline{\xi}^*_i = B$ by \eqref{eq:opt.kkt.a}.
The uniqueness of the solution of \eqref{eq:opt.kkt} can be established by an argument similar to the proof of the monotonicity of $\underline{\xi}^{(j)*}_i$ in $j$.

Next, we show that $(\underline{\xi}^*_i)_{i=1}^\infty$ is optimal and hence $\underline{\xi}^*_i=\xi^*_i$ for all $i\in\pinteger$ by the strict concavity of $\mathcal{R}(x)$.
Note that $\mathcal{R}(x) \le \mathcal{R}(B)$ for $x \in [0,B]$, which yields constant upper bounds on the two infinite sums in \eqref{eq:opt.throughput}.
So by the dominated convergence theorem,
\begin{eqnarray*}
\mathcal{T}_\infty((\underline{\xi}^*_i)_{i=1}^\infty)
&= &\lim_{j\to\infty} \mathcal{T}_\infty((\underline{\xi}^{(j)*}_i)_{i=1}^\infty)\\
&= &\lim_{j\to\infty} \underline{\mathcal{T}}_j((\underline{\xi}^{(j)*}_i)_{i=1}^j)\\
&\ge &\lim_{j\to\infty} \underline{\mathcal{T}}_j((\xi^*_i)_{i=1}^j)\\
&\ge &\lim_{j\to\infty} (\mathcal{T}_\infty((\xi^*_i)_{i=1}^\infty) - \underline{\Delta}_j)\\
&= &\mathcal{T}_\infty((\xi^*_i)_{i=1}^\infty),
\end{eqnarray*}
where
\begin{eqnarray*}
\underline{\Delta}_j
&\eqdef &\sum_{k=j+1}^\infty p(1-p)^{k+w-1} \mathcal{R}(\xi^*_k)\\
& &+ \sum_{k=j+1}^\infty p^2(1-p)^{k+w-1} w \mathcal{R}\left(\frac{B-\sum_{\ell=1}^{k} \xi^*_\ell}{w}\right)\\
&\le &(1-p)^{j+w}\mathcal{R}(B)+pw(1-p)^{j+w}\mathcal{R}(B)
\to 0
\end{eqnarray*}
as $j\to\infty$. Therefore, $(\underline{\xi}^*_i)_{i=1}^\infty$ must coincide with $(\xi^*_i)_{i=1}^\infty$.

Finally, we have
\begin{eqnarray*}[rclqTl]
\xi^*_i
&< &\underline{\xi}^{(i)*}_i
= \frac{B-\sum_{j=1}^i \underline{\xi}^{(i)*}_j}{w} &(Eq.~\eqref{eq:lower.opt.kkt.b})\\
&< &\frac{B-\sum_{j=1}^i \xi^*_j}{w},
\end{eqnarray*}
which further implies
\begin{eqnarray*}
\mathcal{R}'(\xi^*_i)
&\le &p\mathcal{R}'\left(\frac{B-\sum_{j=1}^{i+1} \xi^*_j}{w}\right) + (1-p)\mathcal{R}'(\xi^*_{i+1})\\
&< &\mathcal{R}'(\xi^*_{i+1}),
\end{eqnarray*}
that is, $\xi^*_i > \xi^*_{i+1} = \lim_{j\to\infty} \underline{\xi}^{(j)*}_{i+1}\ge 0$ for all $i\in\pinteger$.
\end{IEEEproof}

\section{Proofs of Results in Section~\ref{sec:upper}}

\begin{IEEEproof}[Proof of Theorem~\ref{th:upper.opt.kkt}]
The uniqueness of the maximizer is an easy consequence of the strict concavity of $\mathcal{R}$.
The KKT conditions for $(\overline{\xi}^{(N)*}_i)_{i=1}^N$ are
\begin{eqnarray*}[l]
p(1-p)^{i+w-1}\mathcal{R}'(\overline{\xi}^{(N)*}_i)\\
\quad - \sum_{k=i}^{N-1} p^2(1-p)^{k+w-1} \mathcal{R}'\left(\frac{B-\sum_{j=1}^k \overline{\xi}^{(N)*}_j}{w}\right)\\
\quad - \sum_{k=w}^\infty p^2(1-p)^{k+N-1} \mathcal{R}'\left(\frac{B-\sum_{j=1}^{N} \overline{\xi}^{(N)*}_j}{k}\right)\\
\quad + \overline{\mu}^{(N)}_i - \overline{\lambda}^{(N)}
= 0,\quad 1\le i\le N,\\
\overline{\mu}^{(N)}_i \overline{\xi}^{(N)*}_i
= 0,\quad 1\le i\le N,\\
\overline{\lambda}^{(N)}\left(B-\sum_{j=1}^N \overline{\xi}^{(N)*}_j\right)
= 0,
\end{eqnarray*}
where $\overline{\mu}^{(N)}_i$ and $\overline{\lambda}^{(N)}$ are all nonnegative.
Equivalently, we have
\begin{subequations}\label{upper.opt.kkt.2}
\begin{eqnarray}[l]
p(1-p)^{i+w-1} \mathcal{R}'(\overline{\xi}^{(N)*}_i) + \overline{\mu}_i^{(N)} \nonumber\\
\quad = p^2(1-p)^{i+w-1}\mathcal{R}'\left(\frac{B-\sum_{j=1}^i \overline{\xi}^{(N)*}_j}{w}\right) \nonumber\\
\qquad + p(1-p)^{i+w}\mathcal{R}'(\overline{\xi}^{(N)*}_{i+1}) + \overline{\mu}_{i+1}^{(N)}, \quad 1\le i<N,\label{eq:upper.opt.kkt.2a}\\
p(1-p)^{N+w-1} \mathcal{R}'(\overline{\xi}^{(N)*}_N) + \overline{\mu}_N^{(N)} \nonumber\\
\quad = \sum_{k=w}^\infty p^2(1-p)^{k+N-1} \mathcal{R}'\left(\frac{B-\sum_{j=1}^{N} \overline{\xi}^{(N)*}_j}{k}\right) + \overline{\lambda}^{(N)}, \nonumber\\
\label{eq:upper.opt.kkt.2b}\\
\overline{\mu}^{(N)}_i \overline{\xi}^{(N)*}_i
= 0,\quad 1\le i\le N,\label{eq:upper.opt.kkt.2c}\\
\overline{\lambda}^{(N)}\left(B-\sum_{j=1}^N \overline{\xi}^{(N)*}_j\right)
= 0\label{eq:upper.opt.kkt.2d}.
\end{eqnarray}
\end{subequations}

We first show that $\sum_{j=1}^N \overline{\xi}^{(N)*}_j < B$.
If $\sum_{j=1}^N \overline{\xi}^{(N)*}_j = B$, then
\begin{eqnarray*}[l]
p(1-p)^{N+w-1} \mathcal{R}'(\overline{\xi}^{(N)*}_N) + \overline{\mu}_N^{(N)}\\
\quad \ge p(1-p)^{N+w-1} \mathcal{R}'(0)
\quad\text{(Eq.~\eqref{eq:upper.opt.kkt.2b})},
\end{eqnarray*}
which implies $\overline{\xi}^{(N)*}_N=0$ because $\mathcal{R}'$ is strictly decreasing.
This further implies  $\sum_{j=1}^{N-1} \overline{\xi}^{(N)*}_j = B$, so
\begin{eqnarray*}[l]
p(1-p)^{N+w-2} \mathcal{R}'(\overline{\xi}^{(N)*}_{N-1}) + \overline{\mu}_{N-1}^{(N)}\\
\quad \ge p(1-p)^{N+w-2}\mathcal{R}'(0)
\quad\text{(Eq.~\eqref{eq:upper.opt.kkt.2a})},
\end{eqnarray*}
and consequently $\overline{\xi}^{(N)*}_{N-1}=0$.
Repeating this backward induction, we finally get $\overline{\xi}^{(N)*}_1=B=0$, which is absurd.
Therefore, $\sum_{j=1}^N \overline{\xi}^{(N)*}_j < B$. Now it can be readily seen that $\overline{\lambda}^{(N)}=0$ (Eq.~\eqref{eq:upper.opt.kkt.2d}) and
\begin{eqnarray*}[l]
p(1-p)^{N+w-1} \mathcal{R}'(\overline{\xi}^{(N)*}_N) + \overline{\mu}_N^{(N)}\\
\quad < p(1-p)^{N+w-1} \mathcal{R}'(0)
\quad\text{(Eq.~\eqref{eq:upper.opt.kkt.2b})},
\end{eqnarray*}
which implies $\overline{\xi}^{(N)*}_N>0$ and $\overline{\mu}_N^{(N)}=0$ (Eq.~\eqref{eq:upper.opt.kkt.2c}).
In view of the fact that
\begin{eqnarray*}
\multicolumn{3}{l}{\mathcal{R}'(\overline{\xi}^{(N)*}_N)}\\
\quad &= &\sum_{k=w}^\infty p(1-p)^{k-w} \mathcal{R}'\left(\frac{B-\sum_{j=1}^{N} \overline{\xi}^{(N)*}_j}{k}\right) \;\text{(Eq.~\eqref{eq:upper.opt.kkt.2b})}\\
&> &\mathcal{R}'\left(\frac{B-\sum_{j=1}^{N} \overline{\xi}^{(N)*}_j}{w}\right),
\end{eqnarray*}
we have
\[
\overline{\xi}^{(N)*}_N
< \frac{B-\sum_{j=1}^{N} \overline{\xi}^{(N)*}_j}{w}.
\]
Furthermore, since
\begin{eqnarray*}[l]
p(1-p)^{N+w-2} \mathcal{R}'(\overline{\xi}^{(N)*}_{N-1}) + \overline{\mu}_{N-1}^{(N)}\\
\quad < p(1-p)^{N+w-2}\mathcal{R}'(0)
\quad\text{(Eq.~\eqref{eq:upper.opt.kkt.2a})},
\end{eqnarray*}
it follows that $\overline{\xi}^{(N)*}_{N-1}>0$ and $\overline{\mu}_{N-1}^{(N)}=0$ (Eq.~\eqref{eq:upper.opt.kkt.2c}), and consequently
\[
\mathcal{R}'(\overline{\xi}^{(N)*}_{N-1})
= p\mathcal{R}'\left(\frac{B-\sum_{j=1}^{N-1} \overline{\xi}^{(N)*}_j}{w}\right) + (1-p)\mathcal{R}'(\overline{\xi}^{(N)*}_N).
\]
This further implies that
\begin{eqnarray*}
\mathcal{R}'(\overline{\xi}^{(N)*}_{N-1})
&< &p\mathcal{R}'\left(\frac{B-\sum_{j=1}^N \overline{\xi}^{(N)*}_j}{w}\right) + (1-p)\mathcal{R}'(\overline{\xi}^{(N)*}_N)\\
&< &\mathcal{R}'(\overline{\xi}^{(N)*}_N)
\end{eqnarray*}
and
\begin{eqnarray*}
\mathcal{R}'(\overline{\xi}^{(N)*}_{N-1})
&> &p\mathcal{R}'\left(\frac{B-\sum_{j=1}^{N-1} \overline{\xi}^{(N)*}_j}{w}\right)\\
& &+ (1-p)\mathcal{R}'\left(\frac{B-\sum_{j=1}^N \overline{\xi}^{(N)*}_j}{w}\right)\\
&> &\mathcal{R}'\left(\frac{B-\sum_{j=1}^{N-1} \overline{\xi}^{(N)*}_j}{w}\right),
\end{eqnarray*}
Thus $\overline{\xi}^{(N)*}_{N-1}>\overline{\xi}^{(N)*}_N$ and
\[
\overline{\xi}^{(N)*}_{N-1}
< \frac{B-\sum_{j=1}^{N-1} \overline{\xi}^{(N)*}_j}{w}.
\]
Repeating such a backward induction finally establishes the theorem.
\end{IEEEproof}

\begin{IEEEproof}[Proof of Theorem~\ref{Theorem:Main-Result2}]
From \eqref{eq:upper.opt.kkt.a}, it is observed that either
\[
\overline{\xi}^{(N)*}_i
< \overline{\xi}^{(N+1)*}_i \quad \text{for all $1\le i\le N$}
\]
or
\[
\overline{\xi}^{(N)*}_i
\ge \overline{\xi}^{(N+1)*}_i \quad \text{for all $1\le i\le N$}.
\]
The latter is however impossible because we would have
\begin{eqnarray*}
\multicolumn{3}{l}{\mathcal{R}'(\overline{\xi}^{(N+1)*}_N)}\\
&= &p\mathcal{R}'\left(\frac{B-\sum_{j=1}^N \overline{\xi}^{(N+1)*}_j}{w}\right)\\
& &+ (1-p)\sum_{k=w}^\infty p(1-p)^{k-w} \mathcal{R}'\left(\frac{B-\sum_{j=1}^{N+1} \overline{\xi}^{(N+1)*}_j}{k}\right)\\
&\ltvar{(a)} &p\mathcal{R}'\left(\frac{B-\sum_{j=1}^N \overline{\xi}^{(N+1)*}_j}{w}\right)\\
& &+ (1-p)\sum_{k=w}^\infty p(1-p)^{k-w} \mathcal{R}'\left(\frac{B-\sum_{j=1}^{N} \overline{\xi}^{(N+1)*}_j}{k+1}\right)\\
&\le &\sum_{k=w}^\infty p(1-p)^{k-w} \mathcal{R}'\left(\frac{B-\sum_{j=1}^{N} \overline{\xi}^{(N)*}_j}{k}\right)\\
&= &\mathcal{R}'(\overline{\xi}^{(N)*}_N),
\end{eqnarray*}
where (a) follows from Lemma~\ref{le:duality.bound} with
\[
\frac{\mathcal{R}'(y)-\mathcal{R}'((x+y)/(k+1))}{\mathcal{R}'(y)-\mathcal{R}'(x/k)}
= 1-\frac{1+\gamma y}{1+k+\gamma(x+y)}.
\]
This shows that $\overline{\xi}^{(j)*}_i$ is strictly increasing in $j$ for $j\ge i$, which implies the existence of $\overline{\xi}^*_i$ and further gives \eqref{eq:opt.kkt.a} by the continuity of $\mathcal{R}'(x)$.
Since $\overline{\xi}^{(j)*}_{i+1}<\overline{\xi}^{(j)*}_i$ for all $1\le i<j$,
\[
\overline{\xi}^*_i
\le \frac{\sum_{j=1}^i \overline{\xi}^*_j}{i}
\le \frac{B}{i}
\to 0
\]
as $i\to\infty$, so we must have $\sum_{i=1}^\infty \overline{\xi}^*_i = B$ by \eqref{eq:opt.kkt.a}.

The uniqueness of the solution of \eqref{eq:opt.kkt} has already been established in Theorem~\ref{Theorem:Main-Result}, and it also implies that
$\overline{\xi}^*_i=\xi^*_i$ for all $i\in\pinteger$.
Here, however, we will give an alternative proof based on the  property of \eqref{eq:upper.opt.def}.
By the dominated convergence theorem,
\begin{eqnarray*}
\mathcal{T}_\infty((\overline{\xi}^*_i)_{i=1}^\infty)
&= &\lim_{j\to\infty} \mathcal{T}_\infty((\overline{\xi}^{(j)*}_i)_{i=1}^\infty)\\
&\ge &\lim_{j\to\infty} (\overline{\mathcal{T}}_j((\overline{\xi}^{(j)*}_i)_{i=1}^j) - \overline{\Delta}_j)\\
&= &\lim_{j\to\infty} \overline{\mathcal{T}}_j((\overline{\xi}^{(j)*}_i)_{i=1}^j)\\
&\ge &\mathcal{T}_\infty((\xi^*_i)_{i=1}^\infty),
\end{eqnarray*}
where
\begin{eqnarray*}
\overline{\Delta}_j
&\eqdef &\sum_{k=j+1}^\infty p(1-p)^{k+w-1} \mathcal{R}(\xi^*_k)\\
& &+ \sum_{k=w+1}^\infty p^2(1-p)^{k+j-1} k \mathcal{R}\left(\frac{B-\sum_{\ell=1}^{j} \xi^*_\ell}{k}\right)\\
&\le &(1-p)^{j+w}\mathcal{R}(B)+p(1-p)^{j+w}\mathcal{R}'(0) B
\to 0
\end{eqnarray*}
as $j\to\infty$. Therefore, $(\overline{\xi}^*_i)_{i=1}^\infty$ must coincide with $(\xi^*_i)_{i=1}^\infty$.
\end{IEEEproof}

\begin{lem}\label{le:duality.bound}
Let $\mathcal{F}$ be a positive, strictly decreasing function on $\nnreal$ and $(a_i)_{i=w}^\infty$ a sequence of positive numbers such that $\sum_{i=w}^\infty a_i=1$.
Let $x, y\ge 0$.
If
\begin{equation}
\mathcal{F}(y)
= \sum_{k=w}^\infty a_k \mathcal{F}\left(\frac{x}{k}\right)\label{eq:duality.bound.condition.1}
\end{equation}
and the sequence
\[
b_k
\eqdef \frac{\mathcal{F}(y)-\mathcal{F}((x+y)/(k+1))}{\mathcal{F}(y)-\mathcal{F}(x/k)}
\]
is strictly increasing in $k$ for all $k=w,w+1,\ldots$ but $k\ne x/y$, then
\[
\sum_{k=w}^\infty a_k \mathcal{F}\left(\frac{x}{k}\right)
< \sum_{k=w}^\infty a_k \mathcal{F}\left(\frac{x+y}{k+1}\right).
\]
\end{lem}

\begin{IEEEproof}
From \eqref{eq:duality.bound.condition.1}, it follows that $\mathcal{F}(y)>\mathcal{F}(x/w)$, and consequently $y<x/w$.
Let $v\eqdef \max\{k\in\pinteger: y<x/k\}$.
Then for $w\le k\le v$, we have $y<x/k$ and
\[
\frac{\mathcal{F}(y)-\mathcal{F}((x+y)/(k+1))}{\mathcal{F}(y)-\mathcal{F}(x/k)}
\le b_v,
\]
so 
\[
\mathcal{F}\left(\frac{x}{k}\right) - \mathcal{F}\left(\frac{x+y}{k+1}\right)
\le (1-b_v)\left(\mathcal{F}\left(\frac{x}{k}\right) - \mathcal{F}(y)\right).
\]
Note that $b_v<\lim_{k\to\infty} b_k = 1$.
On the other hand, for $k > v$ but $k\ne x/y$, we have $y>x/k$ and
\[
\frac{\mathcal{F}(y)-\mathcal{F}((x+y)/(k+1))}{\mathcal{F}(y)-\mathcal{F}(x/k)}
> b_v,
\]
so 
\[
\mathcal{F}\left(\frac{x}{k}\right) - \mathcal{F}\left(\frac{x+y}{k+1}\right)
< (1-b_v)\left(\mathcal{F}\left(\frac{x}{k}\right) - \mathcal{F}(y)\right).
\]
Therefore,
\begin{eqnarray*}
\multicolumn{3}{l}{\sum_{k=w}^\infty a_k \left(\mathcal{F}\left(\frac{x}{k}\right) - \mathcal{F}\left(\frac{x+y}{k+1}\right)\right)}\\
\quad &< &(1-b_v) \sum_{k=w}^\infty a_k \left(\mathcal{F}\left(\frac{x}{k}\right) - \mathcal{F}(y)\right)
= 0.
\end{eqnarray*}
\end{IEEEproof}

\begin{IEEEproof}[Proof of Proposition~\ref{pr:gap}]
\begin{eqnarray*}
\multicolumn{3}{l}{\overline{\Gamma}_N^* - \underline{\Gamma}_N^*}\\
\quad &= &\overline{\mathcal{T}}_N((\overline{\xi}^{(N)*}_i)_{i=1}^N) - \underline{\mathcal{T}}_N((\underline{\xi}^{(N)*}_i)_{i=1}^N)\\
&\le &\overline{\mathcal{T}}_N((\overline{\xi}^{(N)*}_i)_{i=1}^N) - \underline{\mathcal{T}}_N((\overline{\xi}^{(N)*}_i)_{i=1}^N)\\
&\levar{(a)} &\sum_{k=w+1}^\infty p^2(1-p)^{k+N-1} k \mathcal{R}\left(\frac{B-\sum_{j=1}^{N} \overline{\xi}^{(N)*}_j}{k}\right)\\
&\levar{(b)} &\sum_{k=w+1}^\infty p^2(1-p)^{k+N-1} k \mathcal{R}'(0) \frac{B-\sum_{j=1}^{N} \overline{\xi}^{(N)*}_j}{k}\\
&= &p(1-p)^{N+w}\mathcal{R}'(0)\left(B-\sum_{j=1}^N \overline{\xi}^{(N)*}_j\right),
\end{eqnarray*}
where (a) follows from \eqref{eq:lower.opt.def} and \eqref{eq:upper.opt.def}, and (b) follows from the mean value theorem and the concavity of $\mathcal{R}(x)$.
\end{IEEEproof}


\end{document}